\documentclass{article}
\usepackage{cite}
\usepackage{amsmath,amssymb,amsfonts}
\usepackage{algorithmic}
\usepackage{graphicx}
\usepackage{textcomp}
\usepackage{xcolor}
\graphicspath{ {./images/} }
\usepackage{verbatim}

\usepackage{pgfplots}

\usepackage[utf8]{inputenc}
\usepackage{graphicx}    
\usepackage{tikz}
\usepackage{pgfplots}
\usepackage{bchart}
\usetikzlibrary{decorations.pathreplacing}
\usepackage{subfig}
\usepackage{titlesec}
\usepackage{subfig}
\usepackage[numbers,sort&compress]{natbib}
\usepackage{comment}
\usepackage{fancyhdr}
\usepackage{lastpage}
\usepackage{flushend}
\usepackage{wrapfig}
\usepackage{color}
\usepackage{hyperref}
\newcommand{\revise}[1] {{\color{blue} #1}}

\def\BibTeX{{\rm B\kern-.05em{\sc i\kern-.025em b}\kern-.08em
    T\kern-.1667em\lower.7ex\hbox{E}\kern-.125emX}}
    
\title{Serverless Platforms on the Edge: A Performance Analysis}
\author{Hamza Javed,
Adel N. Toosi, %\orcidID{\href{https://orcid.org/0000-0001-5655-5337}{0000-0001-5655-5337}}, 
Mohammad S. Aslanpour %\orcidID{\href{https://orcid.org/0000-0002-1816-6901}{0000-0002-1816-6901}}}
\\
Faculty of Information Technology \\Monash University, Clayton, Australia\\
\{hamza.javed, adel.n.toosi, mohammad.aslanpour\}@monash.edu
}

\begin{document}
\maketitle              % typeset the header of the contribution

\begin{abstract}
The exponential growth of Internet of Things (IoT) has given rise to a new wave of edge computing due to the need to process data on the edge, closer to where it is being produced and attempting to move away from a cloud-centric architecture. This provides its own opportunity to decrease latency and address data privacy concerns along with the ability to reduce public cloud costs. The serverless computing model provides a potential solution with its event-driven architecture to reduce the need for ever-running servers and convert the backend services to an as-used model. This model is an attractive prospect in edge computing environments with varying workloads and limited resources. Furthermore, its setup on the edge of the network promises reduced latency to the edge devices communicating with it and eliminates the need to manage the underlying infrastructure. In this book chapter, first, we introduce the novel concept of serverless edge computing, then, we analyze the performance of multiple serverless platforms, namely, OpenFaaS, AWS Greengrass, Apache OpenWhisk, when set up on the single-board computers (SBCs) on the edge and compare it with public cloud serverless offerings, namely, AWS Lambda and Azure Functions, to deduce the suitability of serverless architectures on the network edge. These serverless platforms are set up on a cluster of Raspberry Pis and we evaluate their performance by simulating different types of edge workloads. The evaluation results show that OpenFaaS achieves the lowest response time on the SBC edge computing infrastructure while serverless cloud offerings are the most reliable with the highest success rate.
\end{abstract}

%\keywords{IoT \and OpenFaaS \and Apache OpenWhisk \and AWS Lambda \and Azure Functions \and AWS Greengrass \and Serverless \and FaaS \and Serverless Edge Computing}
%\end{abstract}
%
%
%
\section{Introduction}
In recent years, we have seen a major increase in new developments in cloud computing due the requirement of handling data at a massive scale. New technologies such as Internet of Things (IoT) and edge devices such as smartphones have created a large influx of data and network traffic that has to be processed at scale.~\cite{shi2016edge}\cite{yu2017survey}. This has given rise to new technologies and computational architectures being introduced into cloud computing in order to maintain the Quality of Service (QoS) of the applications providing these services. In the past few years, deployment architectures have changed from monolithic~\cite{baldini2017serverless} to microservice architectures that aid the developers to build large applications in a scalable manner. Microservice architectures demanded the need of Container-as-a-Service (CaaS)~\cite{hussein2019placement} that provided always-on servers with the ability to scale out easily. However, serverless computing provides a different approach to managing servers in the form of Function-as-a-Service (FaaS)~\cite{castro2017serverless} that proposes the elimination of always-on servers and transition of the services to an on-use basis. 

Serverless architecture uses the approach of using ephemeral containers that can be launched instantaneously on request and be stopped upon completion of computation~\cite{nupponen2020serverless}. This requires the developers to break down their application into multiple functions as the basic units of computation and these functions are hosted in separate containers that are launched upon request and are destroyed when the function execution is over. This architecture provides the opportunity to reduce the resource usage of containers when they are idle and are not receiving requests and also reduces the cost of always-running servers~\cite{nastic2017serverless}~\cite{pinto2018dynamic}. Instead, the cost of the infrastructure is only accumulated for the time the function was actually executed. The serverless computing model removes the need for infrastructure management and this provides a great advantage over monolithic and microservice deployment models.

The serverless computing model provides a great opportunity for edge computing, which consists of an event driven architecture and is subject to varying workloads~\cite{varghese2016challenges}~\cite{premsankar2018edge}. Processing data on the edge of the network provides a solution for limited network bandwidth and latency problems as well as data privacy concerns by executing code closer to devices and end users~\cite{hellerstein2018serverless}. This compute model can provide support for latency-sensitive workloads~\cite{glikson2017deviceless} and help adhere to compliance and privacy regulations. Serverless computing also offers reliability and the ability to scale with no associated management overhead in an edge environment. A serverless architecture for edge computing, what we call it \textit{serverless edge computing}, can provide the benefit of faster deployment, a smaller footprint and increased performance~\cite{baresi2019towards}.

\begin{comment}
The serverless computing model has been adopted by all the major cloud computing providers such as AWS, Azure, GCP, IBM and Alibaba Cloud. Each service provider has their own serverless computing offering. But the cost of adopting a single cloud provider is that this can result in vendor lock-in and can reduce the control the developers have on their applications. This has given rise to open-source serverless platforms that do not require a license to set up and can be set up easily in an on-premises environment. Several studies have been conducted on both public cloud serverless offerings~\cite{malawski2017serverless} and open source serverless platforms~\cite{mohanty2018evaluation} to evaluate the performance and suitability of these platforms when setting up a serverless computing architecture for applications.

Recently, both types of serverless offerings, namely, public cloud and open source, have added support for edge devices to their services. As mentioned earlier, this is 
\end{comment}
\begin{comment}
\revise {Due to the requirement of processing data at the edge of the network, serverless platforms have added support for edge devices to their services. This requires extensive research on whether a serverless architecture can be suitable for an edge environment. There is a current gap in the literature on the suitability of serverless computing on the edge and we aim to provide an analysis in this area with this research.}
\end{comment}

This book chapter focuses on the performance evaluation of serverless platforms of both public cloud and open source contributors in an attempt to analyse the viability of serverless platforms in an edge environment. In order to emulate the edge environment, we aim to set up the serverless platforms on ARM based single-board computers (SBCs), namely, Raspberry Pis and we evaluate the performance of ARM architecture-compatible serverless platforms on these devices. For small-scale infrastructures that are set up in edge environments, single board computers are commonly used. By simulating a similar environment, this study aims to provide insights on the performance of serverless platform on resource constraint devices as this research does not exist in the present literature. The analysed open-source serverless platforms are OpenFaaS,\footnote{OpenFaas (2020), \url{https://docs.openfaas.com}}
%~\cite{openfaas}
Apache OpenWhisk,\footnote{Apache OpenWhisk (2020), \url{https://openwhisk.apache.org/documentation.html}}
%~\cite{openwhisk}, 
AWS Greengrass.\footnote{AWS Greengrass (2020), \url{https://docs.aws.amazon.com/greengrass/}}
%~\cite{awsgreengrass}
We compare their performance with serverless computing services of AWS Lambda\footnote{AWS lambda (2020), \url{https://docs.aws.amazon.com/lambda/index.html}}
%~\cite{awslambda} 
and Azure Functions\footnote{Microsoft Azure Functions (2020), \url{https://docs.microsoft.com/en-us/azure/azure-functions}}
%~\cite{azurefunctions} 
in an attempt to research the suitability of a serverless architecture on the edge. We set up our experiments by configuring these serverless platforms on a cluster of  four Raspberry Pis and subjected them to varying amounts of workloads and analysed the differences in the performance metrics such as response time and success rate for various types of workloads, e.g., CPU-intensive, memory-intensive, and disk-intensive functions.

The organization of this book chapter is as follows: Section~\ref{sec-background} provides a background and motivation on the topic. Section~\ref{sec-related-work} focuses on the related works that have been carried out in this research area and describes serverless computing platforms that we have selected for this study. Section~\ref{sec-performance} discusses our performance evaluation methodology, experimental setup and the analysis of the results. Section~\ref{sec-discussion} presents the discussion on our findings. Finally, Section~\ref{sec-conclustions}  provides a conclusion to our work and outlines future work.

\section{Background}\label{sec-background}
\subsection{Motivation}
Imagine a course-grained application with several dependencies between its services which is deployed on the cloud. The incoming workload is naturally highly dynamic, so the application requires varied amount of computing resources to respond to the dynamism over the time, so called auto-scaling~\cite{aslanUCC}. This scenario applies to a great number of Web applications hosted in the cloud~\cite{AslanpourSim2021}.

The first deployment solution appears resource over-provisioning: providing resources, i.e. VMs, for the application as much as the required resources for the maximum expected incoming workload. This solution will greatly guarantee Quality of Service (QoS), but resource wastage is highly likely. Inefficiency also appears more seriously when hypervisor-based machines, i.e., VMs, with large footprint are intended to host the application. 

Another solution for efficiently deploying monolithic applications is to adopting an auto-scaler to dynamically respond to the workload changes by adding or removing replicas of the application, i.e. adding or removing VMs~\cite{AslanpourSim2021}. This can narrow the gap, but auto-scaling large monolithic applications with several interconnected services will not potentially be a sufficiently smart and quick action. 
It is not smart, as potentially only certain parts of the application may need auto-scaling, not the entire application. A monolithic application is composed of several services tightly coupled with each other and cannot be scaled individually easily. 
The solution is not quick also, as heavyweight hypervisor-based virtualization tends to suffer from long boot-up times for newly provisioned VMs, sacrificing QoS. 
Add to auto-scaling challenges that considerable complexity for deploying auto-scalers will not be practical for non-expert users. 

Given the motivational scenario, this question is raised: ``how we can re-design cloud platforms to properly respond to the dynamic nature of the application?'' Recently, serverless computing has attracted great attention in dealing with such challenges. 
In the following, a historical view of serverless computing will be presented.

\subsection{History of Serverless}
In a nutshell, ``serverless attended the reunion between technological advances such as microservices, containerization, and the idea of effortless auto-scaling and pure pay per use model~\cite{AslanAusPDC}.''
From DevOps perspective, let's fork the advances in development, e.g., coding, and deployment, e.g., installation and maintenance, sides for applications and merge them into a united idea as serverless. 

Development advances: monolithic applications, despite easy deployment, are not easily scaled. Hence, service-oriented architectures for applications appeared and advanced towards today's microservices that reshapes a monolithic application into loosely coupled services that can scale individually. Furthermore, the idea of fully decomposited application arose, that assumes each individual microservice can live by itself, under the name of a function. This led to Function-as-a-Services (FaaS)~\cite{AslanAusPDC} wherein only an individual service is developed for a single unit of task. FaaS also decouples the application for them state and runs stateless functions. Note that the if the state is still required for certain applications while staying in FaaS, the state can be preserved separately by storage services. With this advance, the question is how to take advantage of FaaS in deployment side? and how to bring them in reality?

Deployment advances: given the large footprint of VMs, the containerization~\cite{AslanAusPDC} came into the picture. This means, instead of abstracting the underlying resource from a hypervisor in order for launching an isolated computing resource as VM, one can employ the OS kernel to launch a semi-isolated container. This idea came from small footprint requirements for computing resources to efficiently provision resources for certain applications. With this in mind and the advances in development side, it is obvious that containerization will definitely suit FaaS, where single units of computation are intended to be provisioned and scaled.

One step is left to serverless. 
Such advances helped decoupling monolithic applications to fine-grained FaaS and help provisioning lightweight resources, i.e., containers instead of VMs for making FaaS a reality. However, the challenges of auto-scaling and maintenance are still remained unsolved and poses the question that how to benefit from such advances for efficient execution of applications?

That was a problem until Amazon introduced Lambda~\cite{AslanAusPDC} for running functions in terms of FaaS that run on ephemeral short-lived containers. The auto-scaling and maintenance is left to the serverless platform. Precisely, serverless platforms enable developers to write their applications in terms of functions in any language of interest. Then, the platform is responsible for provisioning resources, e.g., container, per each request (invocation) to the function. Simplistically, once a request is made to the function, the serverless platform will run a new container and after completing the task the container is terminated or stays alive for a short period of time for reusability. Functions are intended to run ephemerally and for short-running tasks. Note that the idea of FaaS and serverless appeared under the same banner historically and occasionally are used interchangeably. An added bonus with serverless offered by cloud providers is also to realize a pure pay pay use pricing model wherein one just pays for the actual execution time of tasks. In conventional cloud pricing models, customers pay for the duration of time the resources are provisioned, regardless of being actually used or not.

\begin{figure}[htpb!]
    \centering
    \includegraphics[width=7cm]{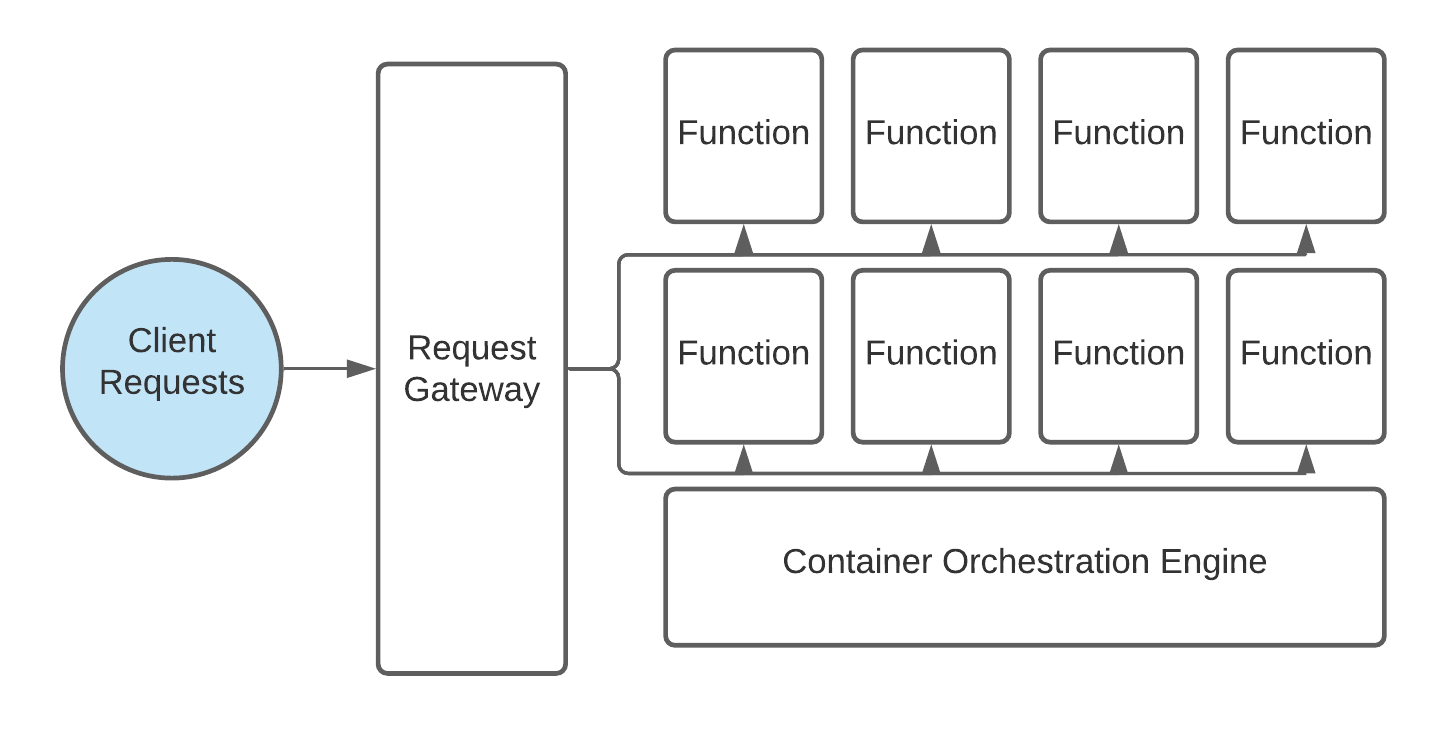}
    \caption{Serverless Architecture}
    \label{fig:serverless}
    \vspace{-0.5cm}
\end{figure}

This provides a huge advantage over other infrastructures such as Platform-as-a-Service (PaaS) or Container-as-a-Service (CaaS) as servers do not need to be running constantly in the background accumulating costs.  Functions start within milliseconds and process individual requests. If there are several simultaneous requests for your function, then the system will create as many copies of the function, that will be managed by the container orchestration system, as needed to meet demand. When demand drops, the application automatically scales down. Dynamic scaling is a benefit of FaaS, and is cost-efficient as well, because providers only charge for the resources that are used, not idle time~\cite{nguyen2019real}. When running on premise, this dynamic nature can also increase platform density, allowing more workloads to run and optimize resource consumption. An event-driven service that needs horizontal scaling can work well as a function, as well as RESTful applications~\cite{chard2019serverless}. Figure~\ref{fig:serverless} shows a schematic serverless architecture.

Given the emergence of AWS Lambda serverless in 2014 as the pioneer, several companies and individual users employed this platform in practice. Other IT companies such as Microsoft (Azure Functions), Google (Google Cloud Functions) IBM (Open Whisk), etc. also attended the market and offered their own serverless platforms. Open-source platforms also came to the picture such as OpenFaaS, Kubeless, etc.

%\begin{comment}

%\end{comment}

\subsection{Serverless at the Edge}
Given the focus area of this research ---serverless edge computing--- this question is raised that ``Why will serverless be good practice for edge computing?''

Edge computing is intended to bring computation closer to data sources and of the biggest stakeholders of edge will be IoT applications~\cite{AslanAusPDC}. 
Let's analyze how IoT applications work. Typically, a group of sensors are located in a particular area, e.g., street, farm, factory, body, etc., in order for collecting and sending data to a computing node for execution.  
The occurrence of such events is highly variable. The execution also would involve short-running tasks such as analyzing if a body temperature is higher than a certain degree or if the moisture content of the soil is at a certain level. 
Add to these characteristics of IoT applications that they will potentially face energy preservation challenges of edge nodes as well. An edge computing platform for several IoT use cases is deemed to perform on low-power devices such as SBCs that demands certain considerations for energy saving.

Obviously, IoT applications with event-driven nature and short-running tasks characteristics will match with serverless platforms functionality. Serverless tends to avoid always-on deployment of applications and instead tends provision containers per requests and to terminate after execution. Hence, this will be a huge step towards energy saving on edge nodes as well.

Theoretically, serverless appears practical for adaptation in edge. This idea has attracted cloud serverless providers to re-design their platform to support edge specific requirements as well. AWS Lambda was once again of the dominant in this move and AWS is now offering relative services such as Greengrass. In academia, researchers also made effort to assess the feasibility of serverless at the edge for different IoT use cases, which will be elaborated in the next section.

\section{Related Works}\label{sec-related-work}

Function-as-a-Service (FaaS) is an event-driven computing execution model that runs in stateless containers and those functions manage server-side logic and state through the use of services~\cite{taibi2020serverless}~\cite{eismann2020serverless}. It allows developers to develop, deploy, and manage those application services as functions without having to maintain their own infrastructure. FaaS provides developers with the flexibility to develop event-driven applications without managing servers. Serverless infrastructure has the ability to scale to zero, i.e. when there are no requests, the servers are stopped and are only running when they are needed~\cite{hall2019execution}.  Serverless computing is a relatively new area and has received a lot of attention in recent years~\cite{van2018serverless}~\cite{lee2018evaluation}. This is due to the potential the serverless computing architecture offers and the need for being validated by research. The work of Van Eyk et al.~\cite{van2018serverless} provides a detailed description and explanation of the serverless computing architecture along with the evolution of cloud computing that led to the rise of serverless computing. Lee et al.~\cite{lee2018evaluation} performed a performance analysis and comparison of the serverless computing services from major cloud providers such as AWS, Azure, GCP and IBM Cloud and analysed the performance of these services in a production environment. %These references need to be added ~\cite{nupponen2020serverless} ~\cite{al2018making} ~\cite{hellerstein2018serverless} ~\cite{hall2019execution} ~\cite{glikson2017deviceless} ~\cite{nastic2017serverless} ~\cite{eismann2020serverless} ~\cite{chard2019serverless} ~\cite{taibi2020serverless} ~\cite{pinto2018dynamic}

The work done by Mohanty et al.~\cite{mohanty2018evaluation} researched multiple open source serverless platforms and comprehensively compared the features and architecture of each platform in order to identify the most suitable platform for production environments. They further elaborated on their research by also comparing the performance of these platforms when running on a Kubernetes cluster. Another study done by Palade et al~\cite{palade2019evaluation} is related to our work as they focus on the performance of open source serverless computing platform at the edge. The serverless platforms that were compared in that study are Kubeless, OpenFaaS, Knative and Apache OpenWhisk.  However, their study does not take into account the computational limitations of edge devices as the platforms are setup on Desktop grade machines and are used to compute the data that is being generated by edge devices. Their study also evaluates qualitative features of these platforms, such as programming language support, container orchestration engine support, monitoring support and CLI interfaces, and the quantitative evaluation, which measures response time, throughput and success rate, and does not include resource intensive tasks. This book chapter provides insights into how a serverless architecture on the edge looks like when we use resource constrained devices such as Raspberry Pis. 

Baresi et al.~\cite{baresi2019towards} discuss the adoption of serverless architectures on the edge and proposes a new serverless platform that can be suitable specifically for an edge environment. Lloyd et al.~\cite{lloyd2018serverless} discusses the performance of AWS Lambda and Azure Functions and compares it across multiple metrics including latency and the effect of warm and cold starts on serverless functions. Shillaker et al.~\cite{shillaker2018provider} evaluates the performance of OpenWhisk and evaluates the response times at varying levels of concurrency. However, none of the works analysed the performance of AWS Lambda function running on the edge using AWS Greengrass and none of the works explicitly compare the performance of both open source and public cloud serverless platforms that is a significant deciding factor in the viability of serverless platforms on the edge. Moreover, to the best of our knowledge, this is the first work that evaluates the performance of open source serverless platforms that are set up on resource constraint edge devices like Raspberry Pis.

In this book chapter, we compare the performance of \textit{AWS Lambda}, \textit{AWS Greengrass} (running local lambda functions), \textit{Azure Functions}, \textit{Apache OpenWhisk}, and \textit{OpenFaaS}. In the following subsections, we briefly describe each of these severless platforms. Table~\ref{tab:qualitative} also provides the qualitative evaluation of the various serverless platforms that we analysed in this book chapter.

%
%
%
%\begin{comment}
\subsection{AWS Lambda}

Amazon Web Services (AWS) offers their serverless computing service called Lambda. AWS Lambda was the pioneer of public cloud serverless computing services as AWS was the first major cloud computing provider to provide a serverless compute service. AWS Lambda follows the FaaS architecture and allows the developers to focus on their application and not worry about managing infrastructure~\cite{villamizar2016infrastructure}. AWS Lambda supports multiple programming languages such as JAVA, Go, Python, Ruby. etc. and the cost model is based the computational expenses of the function execution i.e. the cost is only accumulated for the amount of time that the function is executed. AWS Lambda also provides its Lambda@Edge service which utilizes the AWS Edge Locations in an attempt to reduce latency for its customers and the functions are executed closer to the application which results in lower response times. AWS allows Lambda functions to be run on local edge devices using AWS Greengrass service which is discussed in the next section.
%\end{comment}

%
%
%

\begin{table*}[htpb!]
\scriptsize
\centering
\begin{tabular}{|p{1.6cm}|p{2cm}|p{2cm}|p{2cm}|p{2cm}|p{2cm}|} 
\hline
 & \textbf{OpenFaaS} & \textbf{OpenWhisk} & \textbf{AWS Lambda} & \textbf{AWS Greengrass} & \textbf{Azure Functions} \\ [0.5ex] 
\hline\hline
\textbf{Characteri-stics} & Allows the development of serverless functions in multiple languages and natively supports Docker. Easy deployment. Auto-scaling according to demand. Portable as it runs on existing hardware and can be deployed on both public and private cloud. & Writes functional logic called actions  which can be scheduled and triggered via HTTP. Highly scalable and resilient. & Lambda runs your code on high-availability compute infrastructure and performs all the administration of the compute resources, including server and operating system maintenance. Capacity provisioning and automatic scaling. Code and security patch deployment. Code monitoring and logging. Fault Tolerance. & Runs AWS Lambda functions on the device to respond quickly to local events, interact with local resources, and process data to minimize the cost of transmitting data to the cloud. Support local publish/subscribe messaging between components. Highly secure. & Easily develop and run massively parallel real-time analytics on multiple streams of data – including IoT – using Azure Stream Analytics. With no infrastructure to manage, process data on demand, scale instantly and only pay per job. \\ 
\hline
\textbf{Programm-ing Languages Supported} & Go, NodeJS Python, Java, Ruby, PHP, and C\# & NodeJS, Go, Java, Scala, PHP, Python, Ruby, and Swift  & Java, Go, PowerShell, NodeJS, C\#,
Python, and Ruby & Java, NodeJS, Python & Java, Python, TypeScript, F\#, C\#, Powershell, and JavaScript  \\
\hline
\textbf{Intended Infrastructure} & Kubernetes, Docker Swarm, extendable to other orchestrators, Public Cloud supported & No orchestrator required, Kubernetes supported, Public Cloud supported. & NA & Docker Compose & NA \\
\hline
\textbf{Virtualiz-ation} & Docker & Docker & Micro-VM/firecracker & Micro-VM/firecracker & Docker \\
\hline
\textbf{Triggers} & HTTP, faas-cli & HTTP, OW-CLI & S3, SNS, DynamoDB, CloudWatch, Config Rules, API Gateway, Greengrass. & MQTT Event, AWS Greengrass CLI & Azure Event Hub and Azure Storage, Web Triggering, trigger types and scheduled types. \\
\hline
\textbf{Billing} & NA & NA & Price according to the memory allocation. \$0.00001667 per GB-second. & Monthly Price per device. \$0.16 per month & Price according to the memory allocation. \$0.000016 per GB-second. \\
\hline
\textbf{Limitations} & Does not provide authentication & Open Whisk does not have many options for triggering actions since it is a bit difficult to integrate them. Also, there is a problem in the execution of the concurrent systems. & Maximum function runtime is 15 minutes, and the default timeout is 3 seconds which makes it unsuitable for long-running workloads. The payload for each invocation of a Lambda function is limited to 6MB, and memory is limited to just under 3GB. & Lack of programming language support currently. Steep learning curve. Limited support for integration with other AWS services. & Maximum function runtime is 10 minutes. Memory usage limit is 1.5 GB. This memory is shared among all Functions in the application. \\
\hline
\textbf{Communi-cation Pattern} & Request/Reply & Request/Reply & Request/Reply & Publish/Subscr-ibe and Request/Reply & Request/Reply \\
\hline
\textbf{State} & Stateless & Stateless & Stateless & Stateless & Stateless \\
\hline
\end{tabular}
\vspace{0.3cm}
\caption{Qualitative Evaluation of Serverless platforms}
\label{tab:qualitative}
\end{table*}

%
%
%
%\begin{comment}
\subsection{AWS Greengrass}
AWS Greengrass is a service provided by AWS that allows its customers to run local compute, messaging, and multiple AWS services on a local edge device. AWS Greengrass allows the developer to set up a Greengrass core device that acts as the primary point of connection between the edge setup and the cloud servers. Other devices can either be added as worker nodes for computation or as ‘Things’ that are devices that are able to communicate with the core device. AWS Greengrass primarily uses MQTT~\cite{mqtt} protocol to communicate between the core devices and the IoT things that are registered in the AWS Console. Greengrass is relevant to our study because AWS allows compute services such as AWS Lambda to be run on edge devices in order for the computation to be closer to the application and user. We use ARM devices to set up our serverless platforms. Luckily,  Greengrass natively supports ARM architecture and allows the Greengrass Core to be set up on a Raspberry Pi along with providing the ability to add multiple Raspberry Pi devices as worker nodes for the core device.

\subsection{Azure Functions}
Microsoft Azure provides Azure Functions as its serverless compute service and follows the same paradigm of serverless computing as discussed previously. The underlying infrastructure is managed by Azure Function providing the developer to focus on their code and building their applications following the serverless development paradigm~\cite{kurniawan2019introduction}. Azure Function use authentication to keep out unwanted entities from the system. Azure Functions is a relatively new service, so there are fewer studies that have analysed its performance and compared it to other public cloud serverless offerings.

\subsection{Apache OpenWhisk}
Apache OpenWhisk is an open-source serverless platform currently in incubation. IBM Cloud has built their IBM Cloud Functions on top of Apache OpenWhisk utilizing the power of OpenWhisk’s services. OpenWhisk has multiple components and relatively consumes more resources due to the added components~\cite{quevedo2019evaluating}. For the purposes of this research, we will be utilizing the Lean OpenWhisk branch of the Apache OpenWhisk platform as it is less resource intensive and more suitable for edge environments compared to its full version. Lean OpenWhisk supports the same number of languages as its master branch but is not natively compatible with the ARM architecture. OpenWhisk can be run on Kubernetes or Docker compose~\cite{kuntsevich2018distributed}. For the purposes of this study, we will be setting up Lean OpenWhisk on Docker compose as it consumes less system resources. In order for it to be compatible, custom Docker images of the Lean OpenWhisk platform have to be created for the ARM architecture as it currently only supports x64 and x86 architectures. 

\subsection{OpenFaaS}
OpenFaaS is a lightweight open-source serverless platform that provides native compatibility with the ARM architectures and is easy to set up and start running production workloads. OpenFaaS is compatible with multiple system architectures and can be run on top of Kubernetes and Docker Swarm. OpenFaaS follows the same serverless compute paradigm as the other serverless platforms and provides services for resource and performance monitoring and supports multiple programming such as Python, JavaScript, Ruby etc. Faasd is a variant of OpenFaaS that utilizes Containerd as the container orchestration system and is lighter than OpenFaaS. However, faasd does not support setup on multiple Raspberry Pi devices in a cluster which is why we use OpenFaaS in our experiments.
%\end{comment}
%
%
%

\section{Performance Evaluation}\label{sec-performance}
\begin{figure}[htp]
    \centering
    \includegraphics[width=12cm]{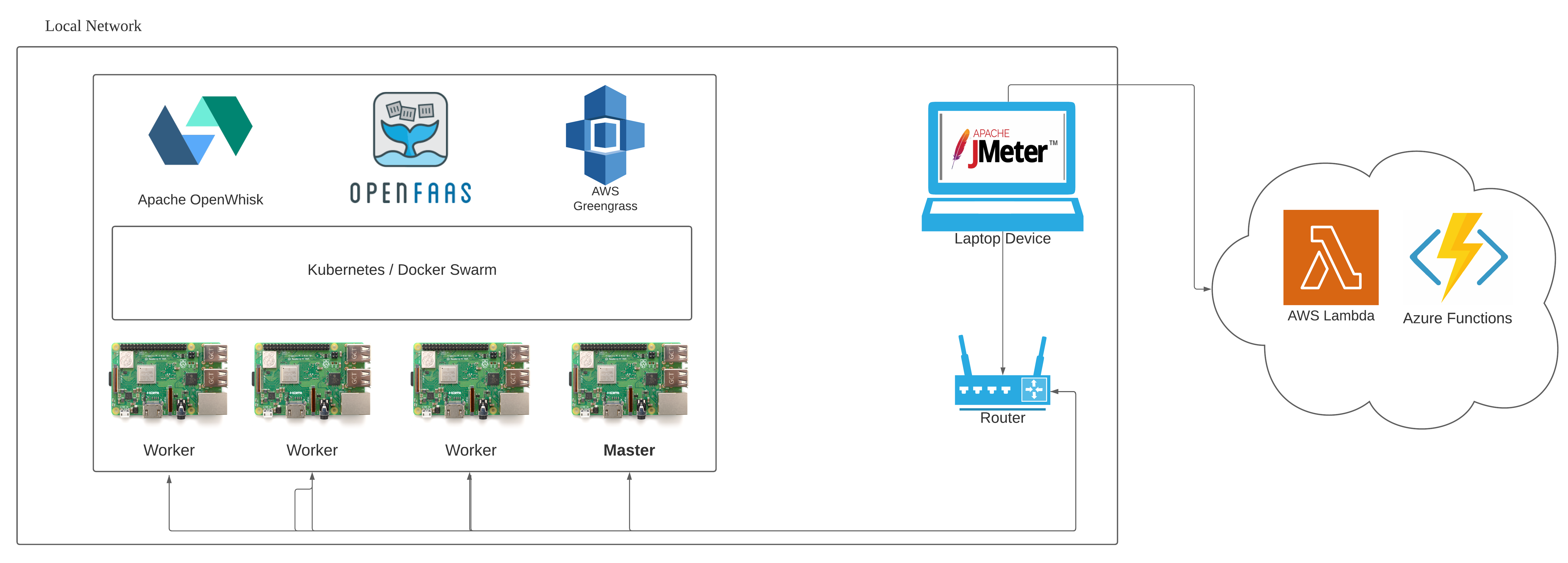}
    \caption{Deployment setup}
    \label{fig:architecture}
    \vspace{-0.6cm}
\end{figure}

We analyze the performance metrics including response times and success rate for the requests measured for each of the aforementioned platforms for the sake of comparison. The response times consists of the time it takes a request to reach the server, the time to execute the function on the server, and the time it takes the response to be delivered back to the client. Success rate measures as the percentage of requests that were executed successfully. As displayed in Figure \ref{fig:architecture}, the local functions are run on a cluster of 4 Raspberry Pis, whereas calls are made to OpenFaas, OpenWhisk and AWS Greengrass in our serverless setup on the local network and also to the cloud functions in AWS Lambda and Azure Functions from the same test machine. This allows us to compare the performance of running local functions as compared to cloud functions which will be a good deciding factor on the viability of running serverless workloads on the edge or running the workloads on the cloud. 
%\begin{comment}
%OpenFaaS natively supports the raspberry pi clusters, whereas OpenWhisk has an intermediary layer of a load balancer in the form of an NGINX server that allows us to distribute the load between the devices. This setup maintains the ability to scale out, allowing us to add more Raspberry Pi devices seamlessly, if we want to increase the computing power according to the load on our setup. However, for the scope of this research, we set the load in a way that it will be sufficiently large to be handled by four Raspberry Pis. Further research can be conducted on more Raspberry Pi devices accordingly.
%\end{comment}

\subsection{Experimental Setup}

We run the experiments on a cluster of four Raspberry Pis, with our requests being generated by a test machine running JMeter on the same local network. For the purpose of accuracy, the experiments are run on an isolated network so that there is no other network overhead that interferes with the network. These tests have been carried in the city of Melbourne, Australia.
The deployment of each platform is done individually in order to make sure no other services are consuming the system resources. This is especially important due to the limited resources available on the edge devices and compute power plays an essential role in our results. We use Raspberry Pi Model 3B+ devices running Raspberry Pi OS which is a Linux distribution designed specifically for Raspberry Pi devices.
\subsubsection{Testbed Setup}
Each Raspberry Pi has a 1.4GHz 64-bit quad-core ARM processor and 1GB RAM. Unless stated otherwise, we used the default deployment settings for each platform, for example, Docker Swarm as the container orchestration system for OpenFaas. \footnote{OpenFaas has switched to Kubernetes as the officially supported orchestration system. This update was provided by OpenFaas after the experiments were conducted in this book chapter.} We use Apache’s Lean OpenWhisk offering that utilizes no Kafka, Zookeeper and no Invokers as separate entities which is suitable for the ARM architecture. The Lean OpenWhisk branch of the Apache OpenWhisk project does not provide compatibility with the ARM architecture of the Raspberry Pi. Each component of OpenWhisk runs as a Docker container and these Docker images have been configured for the ARM architecture in this docker repository.\footnote{Docker (2020), \url{https://docs.docker.com/}} 
%~\cite{docker}.
We created the ARM compatible Docker images to set up OpenWhisk on each Raspberry Pi in the cluster. This setup does not support multiple Raspberry Pi devices in a cluster, so we use an NGINX Load Balancer to distribute our HTTP requests to the Raspberry Pi cluster. 

We use AWS Greengrass to set up our cluster which involves setting up a Greengrass Group on the AWS Management Console and an AWS Greengrass IoT Core on the master Raspberry Pi. The other Raspberry Pi devices are registered as non-core devices and communicate with the core over localhost. The region that the remote Lambda functions are running on Greengrass is set as \textit{ap-southeast-2 i.e. Asia Pacific (Sydney) region}. For our AWS Lambda setup, the region is set as \textit{ap-southeast-2 i.e. Asia Pacific (Sydney, New South Wales) region} in order to maintain consistency and minimize network latency. For our Azure Functions setup, the region is set to \textit{Australia East} which is also the \textit{New South Wales region}. Azure does offer the Australia South East region that is the Victoria region, but for the purposes of comparison with AWS Lambda, the Australia East region is selected. Table~\ref{tab:ping} shows the average ping results from the test machine to the AWS and Azure servers in Sydney and also the ping results to the raspberry pis on the local cluster. This provides us with the time associated with network latency in our experiments.

\begin{table}[h!]
\scriptsize
\centering
\begin{tabular}{|c|c|} 
\hline
\textbf{Destination} & \textbf{Ping} \\ [1.0ex] 
\hline\hline
Raspberry Pi Cluster (local network) & 4ms \\ 
\hline
AWS Sydney Region (ap-southeast-2) & 42ms\\
\hline
Azure Sydney Region (Australia East) & 51ms \\
\hline
\end{tabular}
\caption{Ping results from the test machine}
\label{tab:ping}
\vspace{-1cm}
\end{table}

We use Apache JMeter version 5.3 to generate HTTP requests that invoke the functions deployed on each platform. We run JMeter on a Windows Machine that has a quad-core Core i7 CPU and 8GB RAM. This machine is set up on the same local network as the Raspberry Pi Cluster in order to minimize latency. The JMeter tool is set up to send 1000 requests with different levels of concurrency (5, 10 and 15). These concurrent requests affect the number of simultaneous requests received by each platform. These concurrency levels were decided based on the compute power available to the platforms and each experiment is replicated 3 times in order to maintain statistical accuracy. Specifically for AWS Greengrass, we use the JMeter MQTT plugin which extends the JMeter capability to send MQTT requests instead of HTTP requests as AWS Greengrass requires the user to send MQTT requests.

\subsubsection{Test Functions}

The three types of functions that we use to test our serverless platforms are CPU-intensive function, memory-intensive functions and disk-intensive functions. We use JavaScript for our functions as it is supported by all of our test platforms. The workloads designed for testing these platforms are representative of the various types of workloads that are generated in an edge environment. Due to the lack of research on serverless platforms' performance on resource-constrained devices, this workload has been designed to test how each platform performs when subjected to an emulated workload. In order to simulate a CPU-Intensive function, we multiply two square matrices of 128 by 128 dimensions that has a complexity of O(n\textsuperscript{3}). The execution time of this function varies on the computational power available to the compiler. For the memory-intensive function, we utilize the \textit{memoize()} function of javascript that allows us to store the expensive results of our functions in the cache and retrieve them quickly from the cache if the same result occurs~\cite{berube2007speeding}.

%Below is the code for the memoize function used for this experiment.

\definecolor{RYB2}{RGB}{245,245,245}
\definecolor{RYB1}{RGB}{218,232,252}
\definecolor{RYB4}{RGB}{108,142,191}
\definecolor{RYB3}{RGB}{218,232,100}
\definecolor{RYB5}{RGB}{108,142,100}

\usetikzlibrary{patterns}
\pgfplotsset{compat=newest}
\pgfplotsset{/pgfplots/error bars}

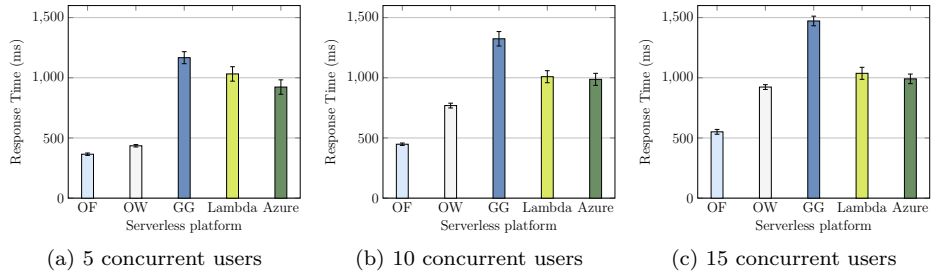
\begin{figure*}
\centering
\subfloat[5  concurrent  users]
{
        \begin{tikzpicture}[thick, scale = 0.45]
        \begin{axis}[
        symbolic x coords={OF,,OW,,GG,,Lambda,,Azure},
        xticklabel style={rotate=0},
        xtick={OF,OW,GG,Lambda, Azure},
        ylabel=Response Time (ms),
        xlabel=Serverless platform,
        label style={font=\large},
        tick label style={font=\large},
        ymin=0, ymax=1600,
        ymajorgrids,
        tick label style={font=\large},
        bar width=10pt,
        ]
        \addplot[ybar,fill=RYB1,error bars/.cd, y dir=both, y explicit] coordinates {(OF,365) +- (30,10)};
        \addplot[ybar,fill=RYB2,error bars/.cd, y dir=both, y explicit] coordinates {(OW,435) +- (30,10)};
        \addplot[ybar,fill=RYB4,error bars/.cd, y dir=both, y explicit] coordinates {(GG,1167) +- (30,50)};
        \addplot[ybar,fill=RYB3,error bars/.cd, y dir=both, y explicit] coordinates {(Lambda,1032) +- (30,60)};
        \addplot[ybar,fill=RYB5,error bars/.cd, y dir=both, y explicit] coordinates {(Azure,923) +- (50,60)};
        \end{axis}
        \end{tikzpicture}%}}
    \label{fig:cpu5}
}
\subfloat[10  concurrent  users]
{
        \begin{tikzpicture}[thick, scale = 0.45]
        \begin{axis}[
        symbolic x coords={OF,,OW,,GG,,Lambda,,Azure},
        xticklabel style={rotate=0},
        xtick={OF,OW,GG,Lambda, Azure},
        ylabel=Response Time (ms),
        xlabel=Serverless platform,
        label style={font=\large},
        tick label style={font=\large},
        ymin=0, ymax=1600,
        ymajorgrids,
        tick label style={font=\large},
        bar width=10pt,
        ]
        \addplot[ybar,fill=RYB1,error bars/.cd, y dir=both, y explicit] coordinates {(OF,448) +- (30,10)};
        \addplot[ybar,fill=RYB2,error bars/.cd, y dir=both, y explicit] coordinates {(OW,769) +- (30,20)};
        \addplot[ybar,fill=RYB4,error bars/.cd, y dir=both, y explicit] coordinates {(GG,1324) +- (30,60)};
        \addplot[ybar,fill=RYB3,error bars/.cd, y dir=both, y explicit] coordinates {(Lambda,1009) +- (30,50)};
        \addplot[ybar,fill=RYB5,error bars/.cd, y dir=both, y explicit] coordinates {(Azure,987) +- (30,50)};
        \end{axis}
    \end{tikzpicture}%}}
     \label{fig:cpu10}
}
\subfloat[15 concurrent users]
{

      \begin{tikzpicture}[thick, scale = 0.45]
        \begin{axis}[
        symbolic x coords={OF,,OW,,GG,,Lambda,,Azure},
        xticklabel style={rotate=0},
        xtick={OF,OW,GG,Lambda, Azure},
        ylabel=Response Time (ms),
        xlabel=Serverless platform,
        label style={font=\large},
        tick label style={font=\large},
        ymin=0, ymax=1600,
        ymajorgrids,
        tick label style={font=\large},
        bar width=10pt,
        ]
        \addplot[ybar,fill=RYB1,error bars/.cd, y dir=both, y explicit] coordinates {(OF,551) +- (30,20)};
        \addplot[ybar,fill=RYB2,error bars/.cd, y dir=both, y explicit] coordinates {(OW,923) +- (30,20)};
        \addplot[ybar,fill=RYB4,error bars/.cd, y dir=both, y explicit] coordinates {(GG,1472) +- (30,40)};
        \addplot[ybar,fill=RYB3,error bars/.cd, y dir=both, y explicit] coordinates {(Lambda,1037) +- (30,50)};
        \addplot[ybar,fill=RYB5,error bars/.cd, y dir=both, y explicit] coordinates {(Azure,991) +- (30,40)};
        \end{axis}
        \end{tikzpicture}%}}

    \label{fig:cpu15}
}
\caption{Median response time with standard error bars on CPU-Intensive tasks for OpenFaas (OF), OpenWhisk (OW), AWS Greengrass (GG), AWS Lambda (Lambda), Azure Functions (Azure) with various number of concurrent users.}
\label{fig:cpu}
\vspace{-0.5cm}
\end{figure*}

%\begin{verbatim}
%function memoizer(func){
%    let memcache = {}
%    return function (n){
%        if (memcache[n] != undefined ) {
%          return memcache[n]
%        } else {
%          let b = func(n)
%          memcache[n] = b
%          return b
%        }
%    }
%}
%\end{verbatim}
To simulate a disk-intensive workload, we created a function to unzip 10 zipped files of size 2.5MB onto the disk. This utilizes both read and write operations for the function. The disk-intensive function requires an extensive setup especially for the cloud functions that we are executing using AWS Lambda and Azure Functions. This is because these services are event-driven and make use of ephemeral containers that are stateless and do not have access to file system that can be shared across all other functions. While a temporary file system for the individual function is included, this ephemeral storage is not intended for durable storage which is why need to attach a shared file system for the functions to access shared data. For AWS Lambda, we created an EFS File System%~\cite{awsefs}
on the AWS Console and provided read/write permissions for Lambda. Then we added the EFS File System in the Lambda configuration to as Local Mount Path%~\cite{awsefs}. 
Azure Functions follows a similar procedure to AWS Lambda. We used Azure Blob Storage to provide file access to our Azure Function. %~\cite{agarwal2012azurebench}. 
For the other platforms, we placed the zip files on each Raspberry Pi in the cluster as the devices do not share a file system. In future work, we would like to test these platforms on actual IoT application workload traces in order to gain better insight on the performance of each serverless platform.

\subsection{Results}

In order to compare the performance of each platform, we measure the response times of each request under different levels of load. %{ In order to minimize the impact of cold starts~\cite{castro2017serverless}, we run the experiment twice back to back without recording so an instance of the function is running when we start our experiment. /} 
The number of concurrent requests directly affects the performance of platforms which alludes to the number of function instances being created in order to handle the requests. The function instances consume the system resources and we want to measure the effect of this on the performance of each platform. Our main aim is to find out the performance issues for each platform and how it handles different types of serverless workloads i.e. CPU-intensive, memory-intensive and disk-intensive functions.
%{ (This paragraph is just sample text)In order to compare the performance of each platform, we measure the response times of each request under different levels of load. In order to minimize the impact of cold starts~\cite{b26}, we run the experiment twice without recording so an instance of the function is running when we start our experiment The number of concurrent requests directly affects the performance of platforms which alludes to the number of function instances being created in order to handle the requests. The function instances consume the system resources and we want to measure the effect this has on the performance of each platform. Our main aim is to find out the performance issues for each platform and how it handles different types of serverless workloads i.e. the type of functions we are running. /}

\begin{figure*}
\centering
\subfloat[5  concurrent  users]
{
        \begin{tikzpicture}[thick, scale = 0.45]
        \begin{axis}[
        symbolic x coords={OF,,OW,,GG,,Lambda,,Azure},
        xticklabel style={rotate=0},
        xtick={OF,OW,GG,Lambda, Azure},
        ylabel=Response Time (ms),
        xlabel=Serverless platform,
        label style={font=\large},
        tick label style={font=\large},
        ymin=0, ymax=1600,
        ymajorgrids,
        tick label style={font=\large},
        bar width=10pt,
        ]
        \addplot[ybar,fill=RYB1,error bars/.cd, y dir=both, y explicit] coordinates {(OF,564) +- (30,10)};
        \addplot[ybar,fill=RYB2,error bars/.cd, y dir=both, y explicit] coordinates {(OW,416) +- (30,20)};
        \addplot[ybar,fill=RYB4,error bars/.cd, y dir=both, y explicit] coordinates {(GG,1425) +- (30,60)};
        \addplot[ybar,fill=RYB3,error bars/.cd, y dir=both, y explicit] coordinates {(Lambda,1083) +- (30,50)};
        \addplot[ybar,fill=RYB5,error bars/.cd, y dir=both, y explicit] coordinates {(Azure,961) +- (30,50)};
        \end{axis}
        \end{tikzpicture}%}}
    \label{fig:mem5}
}
\subfloat[10  concurrent  users]
{
        \begin{tikzpicture}[thick, scale = 0.45]
        \begin{axis}[
        symbolic x coords={OF,,OW,,GG,,Lambda,,Azure},
        xticklabel style={rotate=0},
        xtick={OF,OW,GG,Lambda, Azure},
        ylabel=Response Time (ms),
        xlabel=Serverless platform,
        label style={font=\large},
        tick label style={font=\large},
        ymin=0, ymax=1600,
        ymajorgrids,
        tick label style={font=\large},
        bar width=10pt,
        ]
        \addplot[ybar,fill=RYB1,error bars/.cd, y dir=both, y explicit] coordinates {(OF,639) +- (30,20)};
        \addplot[ybar,fill=RYB2,error bars/.cd, y dir=both, y explicit] coordinates {(OW,506) +- (30,20)};
        \addplot[ybar,fill=RYB4,error bars/.cd, y dir=both, y explicit] coordinates {(GG,1438) +- (30,50)};
        \addplot[ybar,fill=RYB3,error bars/.cd, y dir=both, y explicit] coordinates {(Lambda,1057) +- (30,40)};
        \addplot[ybar,fill=RYB5,error bars/.cd, y dir=both, y explicit] coordinates {(Azure,1076) +- (30,50)};
        \end{axis}
    \end{tikzpicture}%}}
     \label{fig:mem10}
}
\subfloat[15 concurrent users]
{

      \begin{tikzpicture}[thick, scale = 0.45]
        \begin{axis}[
        symbolic x coords={OF,,OW,,GG,,Lambda,,Azure},
        xticklabel style={rotate=0},
        xtick={OF,OW,GG,Lambda, Azure},
        ylabel=Response Time (ms),
        xlabel=Serverless platform,
        label style={font=\large},
        tick label style={font=\large},
        ymin=0, ymax=1600,
        ymajorgrids,
        tick label style={font=\large},
        bar width=10pt,
        ]
        \addplot[ybar,fill=RYB1,error bars/.cd, y dir=both, y explicit] coordinates {(OF,694) +- (30,20)};
        \addplot[ybar,fill=RYB2,error bars/.cd, y dir=both, y explicit] coordinates {(OW,621) +- (30,20)};
        \addplot[ybar,fill=RYB4,error bars/.cd, y dir=both, y explicit] coordinates {(GG,1508) +- (30,40)};
        \addplot[ybar,fill=RYB3,error bars/.cd, y dir=both, y explicit] coordinates {(Lambda,1029) +- (30,50)};
        \addplot[ybar,fill=RYB5,error bars/.cd, y dir=both, y explicit] coordinates {(Azure,1026) +- (30,40)};
        \end{axis}
        \end{tikzpicture}%}}

    \label{fig:mem15}
}
\caption{Median response time with standard error bars on Memory-Intensive tasks for OpenFaas (OF), OpenWhisk (OW), AWS Greengrass (GG), AWS Lambda (Lambda), Azure Functions (Azure) with various number of concurrent users.}
\label{fig:mem}
\vspace{-0.5cm}
\end{figure*}
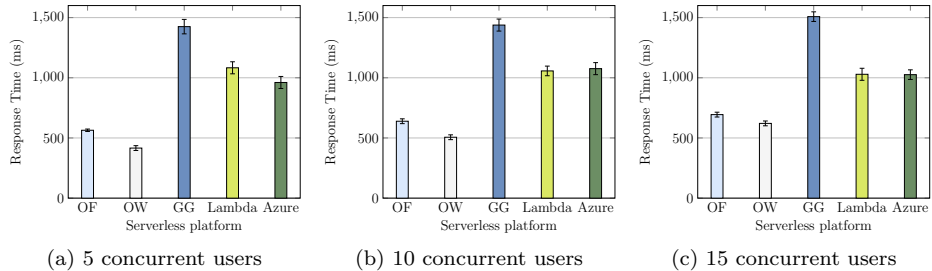

\subsubsection{CPU-Intensive Functions}

Figure~\ref{fig:cpu} shows the median response times for each platform for the 1000 requests that we sent under different concurrency. These response times include the time taken to send the request, the time taken to execute the function and the time taken to receive the result to our test machine. This is an asynchronous request so it waits for the function to be executed and then receives the response. In each of the results, the lowest response time recorded is for the OpenFaaS platform. We observe that the lowest response time is when we send 5 concurrent requests to OpenFaaS with values around 400ms. OpenWhisk performs fairly similarly with values around 450ms. These response times increase as we increase the number of concurrent requests to 10 where OpenWhisk is impacted considerably more than OpenFaaS as the response time increases to 750ms. OpenFaaS response time rises to 448ms when dealing with 10 concurrent requests. The response times for OpenWhisk rise even more when subject to 15 concurrent requests as the response time is 923ms. As compared to OpenFaaS, the response time remain fairly stable when we increase the concurrent requests. 

\begin{table*}[htpb!]
\scriptsize
\centering
\begin{tabular}{|c |c |c |c |c |c|} 
\hline
 & \textbf{OpenFaaS} & \textbf{OpenWhisk} & \textbf{AWS Greengrass} & \textbf{AWS Lambda} & \textbf{Azure Functions} \\ [0.5ex] 
\hline\hline
CPU & 100\% & 99.1\% & 100\% & 100\% & 100\% \\ 
\hline
Memory & 99.5\% & 99.3\% & 100\% & 100\% & 100\% \\
\hline
Disk & 99.9\% & 99.6\% & 100\% & 100\% & 100\% \\
\hline
\end{tabular}
\caption{Request Success Rate for platforms under different workloads}
\label{tab:success}
\vspace{-1cm}
\end{table*}

As shown in Table~\ref{tab:success}, OpenWhisk records the worst success rate in our experiments for CPU-intensive workloads. We recorded 8 failed requests out of 1000 as a result of an increase in the system's resource consumption. OpenFaaS records a perfect success rate with no failed requests deducing that it handles CPU-intensive workloads very well under various loads. These results show that performance of OpenWhisk decreases significantly as we increase the number of concurrent requests and the reason for that is because OpenWhisk consumes more compute power to run its components as compared to OpenFaaS which is more lightweight. If we analyze the response times for AWS Greengrass, we can compare the rise in latency is not that significant when we increase the number of concurrent requests accordingly. However, AWS Greengrass records the highest response times out of all the platforms, due to the time it takes for the request to reach the AWS server to trigger the function on the core device. AWS Lambda and Azure Functions performs very similarly for each level of concurrency and show very slight changes in latency which is predictable as these platforms do not have as limited resources as compared to the other platforms and provide highest level of scalability. For our CPU-intensive functions, AWS Greengrass, AWS Lambda and Azure Functions and OpenFaas do not have a single failed request as compared to OpenWhisk.

\subsubsection{Memory-Intensive Functions}

Figure~\ref{fig:mem} demonstrates the results for the performance of each serverless platform for memory-intensive functions. The response times were recorded for each platform until varying concurrent requests in order to observe the effect it has on the performance. The lowest response times, in each of the tests for memory-intensive functions, were for Apache OpenWhisk that performed considerably better than the other platforms and the results were quite stable with only a gradual increase in response time when we increased the number of concurrent requests. The lowest median response time for OpenWhisk is 416ms for 5 concurrent requests and it reaches a maximum of 621ms for 15 concurrent requests. OpenFaaS, however, demonstrated higher response times than OpenWhisk for memory-intensive functions but the response time varied less between different concurrent requests which proposes that this increase in latency is less likely to be caused by OpenFaaS ability to scale and is more likely attributed to memory management for function execution in the platform.

The highest response times were for AWS Greengrass that maintained its values between 1400ms to 1500ms between varying concurrent requests. We observed that AWS Greengrass took more time for memory-intensive functions than CPU-intensive functions but is affected less by change in concurrent results. In fact, there is only a 13ms difference between the results we send 5 concurrent requests and 10 concurrent requests and Greengrass only witnesses a slight increase in latency when we increase the concurrent requests to 15. Both AWS Lambda and Azure Functions perform very similarly with little changes in latency when we increase the number of concurrent requests. Both of the platforms provided similar results to our CPU-intensive tasks. This is attributed to both of the cloud serverless offerings’ ability to scale indefinitely and also due to the higher computational power available to the functions. However, as the functions are not being executed locally and need to communicate with the cloud servers, the latency is much higher as compared to the platforms that are set up locally.
\vspace{-0.2cm}
\subsubsection{Disk-Intensive Functions}

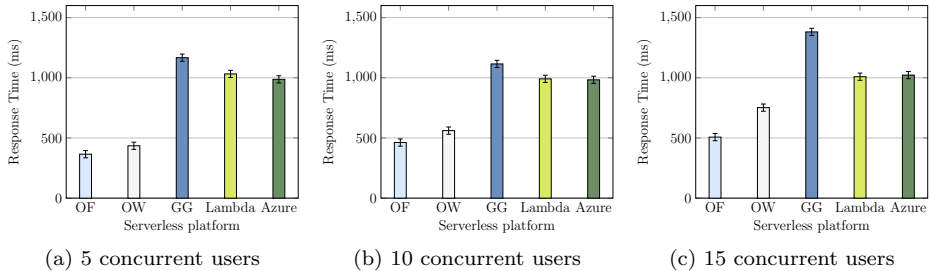
\begin{figure*}
\centering
\subfloat[5  concurrent  users]
{
        \begin{tikzpicture}[thick, scale = 0.45]
        \begin{axis}[
        symbolic x coords={OF,,OW,,GG,,Lambda,,Azure},
        xticklabel style={rotate=0},
        xtick={OF,OW,GG,Lambda, Azure},
        ylabel=Response Time (ms),
        xlabel=Serverless platform,
        label style={font=\large},
        tick label style={font=\large},
        ymin=0, ymax=1600,
        ymajorgrids,
        tick label style={font=\large},
        bar width=10pt,
        ]
        \addplot[ybar,fill=RYB1,error bars/.cd, y dir=both, y explicit] coordinates {(OF,365) +- (30,30)};
        \addplot[ybar,fill=RYB2,error bars/.cd, y dir=both, y explicit] coordinates {(OW,435) +- (30,30)};
        \addplot[ybar,fill=RYB4,error bars/.cd, y dir=both, y explicit] coordinates {(GG,1167) +- (30,30)};
        \addplot[ybar,fill=RYB3,error bars/.cd, y dir=both, y explicit] coordinates {(Lambda,1032) +- (30,30)};
        \addplot[ybar,fill=RYB5,error bars/.cd, y dir=both, y explicit] coordinates {(Azure,987) +- (30,30)};
        \end{axis}
        \end{tikzpicture}%}}
    \label{fig:disk5}
}
\subfloat[10  concurrent  users]
{
        \begin{tikzpicture}[thick, scale = 0.45]
        \begin{axis}[
        symbolic x coords={OF,,OW,,GG,,Lambda,,Azure},
        xticklabel style={rotate=0},
        xtick={OF,OW,GG,Lambda, Azure},
        ylabel=Response Time (ms),
        xlabel=Serverless platform,
        label style={font=\large},
        tick label style={font=\large},
        ymin=0, ymax=1600,
        ymajorgrids,
        tick label style={font=\large},
        bar width=10pt,
        ]
        \addplot[ybar,fill=RYB1,error bars/.cd, y dir=both, y explicit] coordinates {(OF,462) +- (30,30)};
        \addplot[ybar,fill=RYB2,error bars/.cd, y dir=both, y explicit] coordinates {(OW,561) +- (30,30)};
        \addplot[ybar,fill=RYB4,error bars/.cd, y dir=both, y explicit] coordinates {(GG,1115) +- (30,30)};
        \addplot[ybar,fill=RYB3,error bars/.cd, y dir=both, y explicit] coordinates {(Lambda,991) +- (30,30)};
        \addplot[ybar,fill=RYB5,error bars/.cd, y dir=both, y explicit] coordinates {(Azure,983) +- (30,30)};
        \end{axis}
    \end{tikzpicture}%}}
     \label{fig:disk10}
}
\subfloat[15 concurrent users]
{

      \begin{tikzpicture}[thick, scale = 0.45]
        \begin{axis}[
        symbolic x coords={OF,,OW,,GG,,Lambda,,Azure},
        xticklabel style={rotate=0},
        xtick={OF,OW,GG,Lambda, Azure},
        ylabel=Response Time (ms),
        xlabel=Serverless platform,
        label style={font=\large},
        tick label style={font=\large},
        ymin=0, ymax=1600,
        ymajorgrids,
        tick label style={font=\large},
        bar width=10pt,
        ]
        \addplot[ybar,fill=RYB1,error bars/.cd, y dir=both, y explicit] coordinates {(OF,507) +- (30,30)};
        \addplot[ybar,fill=RYB2,error bars/.cd, y dir=both, y explicit] coordinates {(OW,752) +- (30,30)};
        \addplot[ybar,fill=RYB4,error bars/.cd, y dir=both, y explicit] coordinates {(GG,1381) +- (30,30)};
        \addplot[ybar,fill=RYB3,error bars/.cd, y dir=both, y explicit] coordinates {(Lambda,1009) +- (30,30)};
        \addplot[ybar,fill=RYB5,error bars/.cd, y dir=both, y explicit] coordinates {(Azure,1022) +- (30,30)};
        \end{axis}
        \end{tikzpicture}%}}

    \label{fig:disk15}
}
\caption{Median response time with standard error bars on Disk-Intensive tasks for OpenFaas (OF), OpenWhisk (OW), AWS Greengrass (GG), AWS Lambda (Lambda), Azure Functions (Azure) with various number of concurrent users.}
\label{fig:disk}
\vspace{-0.5cm}
\end{figure*}

Figure~\ref{fig:disk} shows the results for the evaluation of each platform for disk-intensive functions. The lowest response times for this experiment are recorded by OpenFaaS and the platform scaling very well when we increased the number of concurrent requests. The response time only increased by 50ms between 10 and 15 concurrent requests which demonstrates the new functional replicas were able to handle the increased load easily. However, we noticed that OpenWhisk recorded higher response times as compared to OpenFaaS and when we increased the load from 10 to 15 concurrent requests, it caused an increase in response time of 200ms. The success rates for OpenFaaS and OpenWhisk were also very acceptable as we recorded very few failed requests as compared to the other function types. AWS Greengrass showed similar trends to OpenWhisk as the response times for 5 and 10 concurrent requests are very similar but an increase in load to 15 requests caused the latency to increase by 220ms. This can be attributed to the creation of new function replicas in both cases of OpenWhisk and AWS Greengrass. AWS Lambda and Azure Functions both displayed similar results for this experiment, as we did not notice any changes in latency caused due to an increase in load or the function type. The file storages for both platforms were configured on the same server as the functions, which shows us that there is minimum latency for disk read and write operations. The results for the success rates for AWS Greengrass, AWS Lambda and Azure Functions were the same as the previous experiments as they recorded perfect success rates with no failed requests.

\vspace{-0.5cm}

\section{Discussion}\label{sec-discussion}
Our experimental results show that OpenFaaS performs considerably better on the edge than the other platforms. We can attribute this to the lean architecture of the platform which suits very favorably to edge devices and consumes fewer resources which is very suitable for edge devices. Along with that, it provides native support for the ARM architecture, along with the ability to scale easily by adding more raspberry pi devices. %It is an open-source project and receives constant updates from its community. 
However, the results vary for high loads especially for CPU-intensive and memory-intensive workloads. We found out that the performance of Apache OpenWhisk is inferior to OpenFaaS in our setup as OpenWhisk does not provide full support for the ARM architecture and the lean version restricts its performance when handling increased workloads. This performance bottleneck may be attributed the high memory usage of the platform that leads to resource starvation for new containers. Furthermore, the ability to add more worker nodes with ease is an integral part of edge computing setup, which OpenWhisk does not support natively.

AWS Greengrass performed very consistently across our experiments, but the increased response times are a considerable factor when making the decision to deploy this platform on the edge. The cause of these increased response times is the added latency of the requested being routed to the AWS cloud servers. The increased latency for CPU-intensive tasks also suggests that the functions take longer to compute due to resource starvation. AWS Greengrass seamlessly extends AWS to edge devices so they can act locally on the data they generate, while still using the cloud for management, analytics, and durable storage. It enables the connected devices to run AWS Lambda functions, keep device data in sync, and communicate with other devices securely. AWS Greengrass provides support for ARM architecture and provides the ability to scale out by adding more worker devices. AWS Greengrass provides an opportunity for further research to be conducted by testing its performance in an isolated edge without connectivity to the cloud. 

If we want consistent performance regardless of the amount of workload, the cloud serverless offerings of AWS Lambda and Azure Functions provide a stable service and can scale according to the increased load. This demonstrates that for workloads that require increased computational capacity, along with the ability to handles a workloads at an increased scale, the cloud serverless offerings are very suitable. However, the overall latency to cloud is higher than the edge devices as experiments show and running serverless platforms such as OpenFaaS on the edge provides promising prospects to meet the stringent latency requirements of emerging real-time applications such as autonomous vehicles. Nevertheless, our work provides the groundwork for comprehensively benchmarking the performance of serverless platforms on the edge, specifically on devices with ARM architecture.
\vspace{-0.5cm}
\section{Conclusions and Future Work}\label{sec-conclusions}

Performing serverless computations on the edge proposes numerous advantages in an event-driven architecture, allowing us to generate more data on the edge without having to worry about managing its applications. This book chapter analysed the performance of serverless platforms on the edge, specifically on ARM architecture edge devices such as Raspberry Pis. We performed comprehensive performance tests on the compatible serverless platforms of OpenFaaS, Apache OpenWhisk and AWS Lambda functions running locally in the AWS Greengrass platform on edge devices, and compared their performance to cloud serverless offerings of AWS Lambda and Azure Functions to research the viability of setting up a serverless architecture in an edge environment. Based on that, the results demonstrated that OpenFaaS is the most suitable platform for an edge setup as it provides a lightweight architecture with support for simple and rapid scaling. We used the metrics of response time and success rate for each platform to compare the performance on how each platform would cope in an edge setup. AWS Greengrass provides a promising opportunity in this environment for further research due to its native support for ARM architecture and its development support from AWS. For future research, we aim to analyse other suitable serverless platforms %such as X and Y 
and develop ARM architecture support and further research the viability of serverless architecture on the edge.
%
%
%

\bibliographystyle{plain}
\bibliography{bibliography.bib}

\end{document}